\documentclass{iauc}
\usepackage{graphicx}

\def\bmk{B$-$K}
\def\ebmv{E(B$-$V)}
\def\lya{Ly$\alpha$}
\def\nhi{$N$(H~I)}

\pubyear{2005}
\volume{199}
\pagerange{1--}
\jname{Probing Galaxies through Quasar Absorption Lines}
\editors{P. R. Williams, C. Shu, and B. M\'{e}nard, eds.}

\title[Absorbers In Radio QSO Surveys] 
{Absorption Systems In Radio-Selected QSO Surveys}

\author[Ellison]   
{Sara L. Ellison$^1$}

\affiliation{$^1$University of Victoria, Dept. Physics \& Astronomy,
Victoria, BC, V8P 1A1, Canada \break 
email: sarae@uvic.ca}

\begin{document}

\maketitle

\begin{abstract}
Radio-selected samples of quasars with complete optical identifications
offer an ideal dataset with which to investigate dust bias associated
with intervening absorption systems.  Here, we review our work
on the Complete Optical and Radio Absorption Line System (CORALS)
survey whose aim is to quantify this bias and assess the impact
of dust on absorber statistics.  First, we review previously 
published results on the number
density and gas content of high column density absorbers over the
redshift range $0.6 < z < 3.5$.  We then present the latest results
from CORALS which focus on measuring the metal content of our
unbiased absorber sample and an investigation of their optical--IR
colours.  Overall we find that although dust is unarguably present
in absorption galaxies, the level appears to be low enough that the statistics
of previous magnitude limited samples have not been severely affected
and that the subsequent reddening of background QSOs is small.


\end{abstract}

\section{Introduction}

The issue of dust bias in QSO absorption line surveys is as
old as the study of damped Lyman alpha systems (DLAs) themselves, dating
back some two decades to the mid 1980s (e.g. Ostriker \& Heisler
1984).  However, it is evident from the number of contributions 
at this conference that the dust content of DLAs and its 
impact on our observations remains a debated and relevant
issue.  The question can be simply stated: is there a sizable 
population of DLAs with enough dust to dim background QSOs and cause 
them to drop out of magnitude limited samples?  This question is
pivotal for the interpretation of DLA statistics since QSO absorption
lines are usually heralded as a 'representative probe' of the gas in
the high redshift universe.  This is true, in the sense that we do not
rely on starlight to detect DLA galaxies.  However,
if our observations are biased against chemically evolved systems with
high dust fractions, the lore that underpins the
utility of DLAs as early galaxy probes will have to be
revised.  Under various assumptions
of the chemical composition and dust properties of high redshift
galaxies, a number of theoretical models have advocated dust-induced
selection effects (e.g. Ostriker \& Heisler 1984; Fall \& Pei 1993; 
Churches et al. 2004; Vladilo \& P\'{e}roux 2005).  The predicted 
impact on observations is many-fold
and ranges from under-estimates of the absorber number density and
total neutral gas density, to the metallicities inferred from 
volatile elements such as Zn.  In addition to this theoretical
impetus, it is tempting to appeal to dust bias in order to
explain a number of observational trends,
such as the anti-correlation between N(HI) and [Zn/H] (Boiss\'{e} et al.
1998) and the very mild evolution of the mean metallicity over 
$0.5 < z < 3.5$ (e.g. Kulkarni et al. 2005).  

In order to quantify the effect of dust induced biases in previous
surveys for DLAs, it is necessary to construct a sample of QSOs
that is optically complete and has no magnitude limit.  One
way to achieve this is to select QSOs at wavelengths that are
not affected by the presence of dust, such as in the radio
or X-ray regimes, and then conduct deep optical observations 
to identify their counterparts.  Any survey for intervening
absorbers based on such a sample will be free from extinction
induced biases and its statistics can be compared with magnitude
limited samples.  This simple strategy has formed the crux of
the Complete Optical and Radio Absorption Line System (CORALS) 
survey for absorbers in the radio-selected Parkes 0.25
Jy (PQJ) QSO sample of Jackson et al. (2002).  Follow-up
optical observations (combined with position matching in
existing digitised surveys) resulted in optical counterparts
for every QSO in the PQJ sample, making it an ideal sample
with which to investigate dust bias.

\section{Results From CORALS}

\subsection{High Redshift DLAs - Number Density and Neutral Gas Content at $1.8 < z < 3.5$}\label{highz}


The first results from CORALS were published by Ellison et al. (2001);
we only briefly review the main results of that work here and reserve the
majority of this proceedings for newer results and a discussion of
the survey's implications.   In short, Ellison et al. (2001) 
conducted a search for DLAs with $z_{\rm abs} > 1.8$ towards
the 66 $z_{\rm em} > 2.2$ QSOs in the PQJ sample, covering
a total redshift path of $\Delta z \sim 55$.
The number density of DLAs
per unit redshift ($n(z)$) and the neutral gas mass density (expressed
as a fraction of the closure density, $\Omega_{\rm DLA}$) were
both found to be consistent with previous surveys, indicating
that dust is not a significant bias in optically selected samples.
However, Ellison et al. (2001) also tentatively pointed out a
mild dependence of $n(z)$ and $\Omega_{\rm DLA}$ on QSO magnitude,
although this dependence seemed to flatten at magnitudes fainter
than $B \sim 19.5$ (see also the poster by Smette et al. at this
meeting).  In Ellison et al. (2004) we speculated that this may
be due to a combination of a small amount of extinction
(i.e. not enough to affect the statistics of the sample as a whole)
and the shape of the QSO optical luminosity function.

\subsection{Low Redshift MgII Systems - Inferences on DLA Number Density at
$0.6<z<1.7$}\label{lowz}

\begin{figure}
\centerline{\rotatebox{0}{\resizebox{8cm}{!}
{\includegraphics{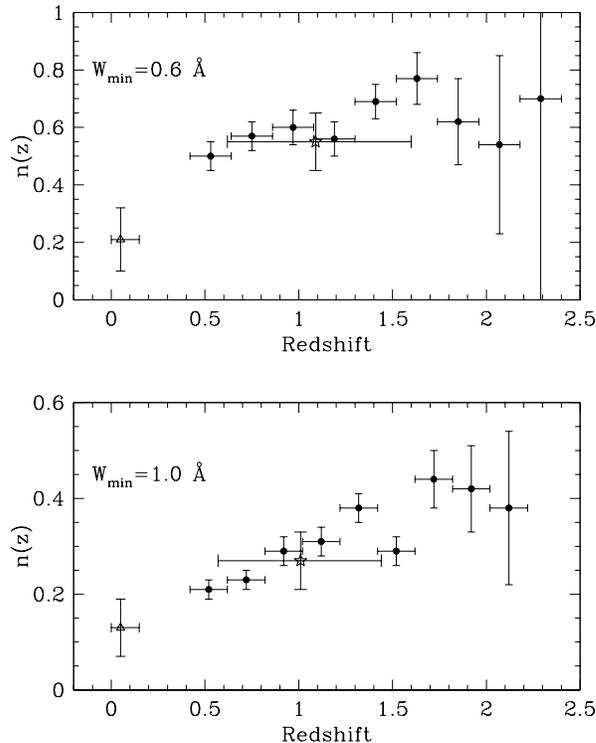}}}}
  \caption{Number density of CORALS MgII absorbers (open stars) for two EW
cuts compared with the SDSS-based survey Nestor et al. (2005, filled 
circles).  The low redshift point is taken from Churchill (2001).}\label{nz}
\end{figure}

Although the majority of known DLAs are at $z>1.7$, this is not
where we might expect dust bias to be most severe.  It is true
that galaxies with large dust masses have been detected at $z>2$
with sub-mm instruments such as SCUBA.  However, it seems unlikely
that low mass stars, the usual culprits for dust formation at
low $z$, can have significantly contributed to high
redshift galaxies.  By $z<1.5$, on the other hand, active star
formation has been on-going at high rates for some 4 billion
years, so we may expect that galaxies have accrued significantly
more dust by these epochs.

Conducting a low redshift survey analogous to the high redshift
campaign described in the previous section is impractical due to
HST's small aperture (and now its complete lack of a UV spectrograph).
However, much can be learned from a ground based survey of strong
metal line systems.  As Sandhya Rao has described at this meeting,
targeting absorbers with high rest equivalent widths (EW $> 0.5$ \AA)
of MgII $\lambda 2796$ and FeII $\lambda 2600$ is an efficient
way of pre-selecting DLA candidates (see also Rao \& Turnshek 2000).
We have therefore searched for strong MgII systems towards our
sample of radio-selected QSOs over the redshift range $0.6 < z_{\rm abs} < 1.7$.
Our sample is slightly different from the high $z$ sample described in
the previous section, in that we observed QSOs in the range
$1.80 < z_{\rm em} < 2.55$.  This modification has been made so that
we optimise our redshift coverage of MgII systems which will lie
redwards of the QSO's \lya\ emission.  Out of 75 QSOs we have
identified 47 MgII absorbers with EW $>$ 0.3 \AA.
We compare the number density of MgII systems in our low
redshift CORALS survey to the SDSS sample of Nestor, Turnshek
\& Rao (2005) in Figure \ref{nz}.  The agreement is excellent,
indicating that our complete survey has not uncovered a large
population of MgII absorbers that have been previously overlooked
by magnitude limited samples.

\begin{figure}
\centerline{\rotatebox{270}{\resizebox{8cm}{!}
{\includegraphics{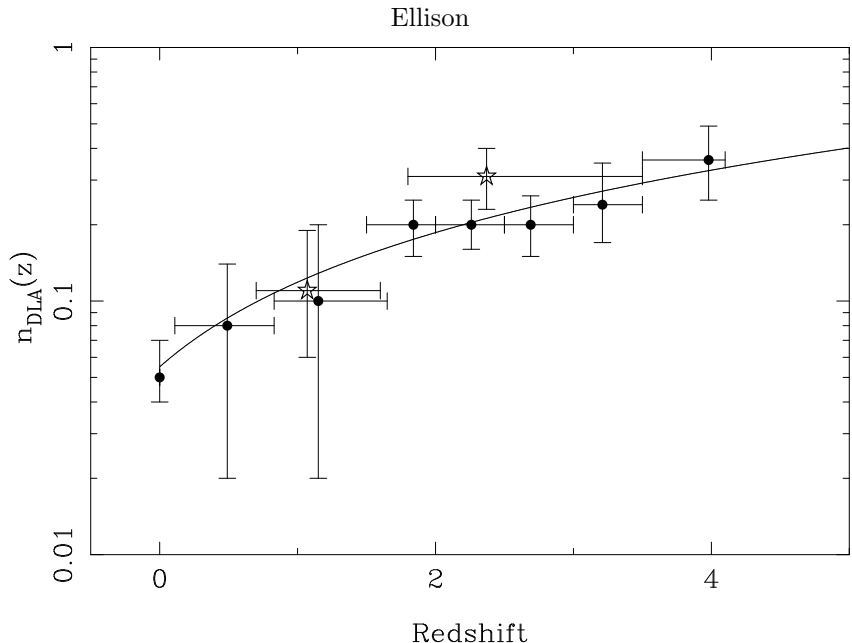}}}}
  \caption{Number density of DLAs from CORALS (open stars):
at high redshift from Ellison et al.
(2001) and low redshift, as inferred from MgII absorber statistics
(Ellison et al. 2004) with the revised DLA confirmation rate of 35\%
(Rao, this meeting).  Solid points show results of the optical
compilation of Storrie-Lombardi \& Wolfe (2000).}\label{nz_dla}
\end{figure}

In Ellison et al. (2004) we assumed that 50\% of high EW MgII and FeII 
absorbers would be DLAs (Rao \& Turnshek 2000).  At this meeting, 
Sandhya Rao has reported improved statistics
from their low redshift DLA survey and have revised the 50\%
confirmation rate of DLAs from strong metal line systems down
to 35\%.  The implied number density of low redshift DLAs in our MgII
sample using this new confirmation rate
(and the high redshift $n(z)$ determined by Ellison et al 2001)
is shown in Figure \ref{nz_dla} and is in excellent agreement with
the compilation of Storrie-Lombardi \& Wolfe (2000).

\subsection{High Redshift DLAs - Metallicity}

In the introduction I gave two examples of observational evidence
that DLAs may be caused by a dust obscuration bias; both were
based on metallicity measurements.  The obvious next step is
therefore to obtain high resolution spectra of the CORALS DLAs,
in order to determine element abundances.  We have recently
completed this work (Akerman et al. 2005) using
the echelle spectrographs on the VLT
(UVES), Keck (ESI) and Magellan (MIKE).  We have measured
[Zn/H] (or obtained upper limits) for 20/22 DLAs in our high $z$
sample; the last two DLAs are towards a single QSO that is
simply too faint even for 8--10-m class telescopes.  However,
the relatively low N(HI) in these systems means that it is
extremely unlikely that these missed absorbers will change our
column density weighted statistics.

\begin{figure}
\centerline{\rotatebox{270}{\resizebox{9cm}{!}
{\includegraphics{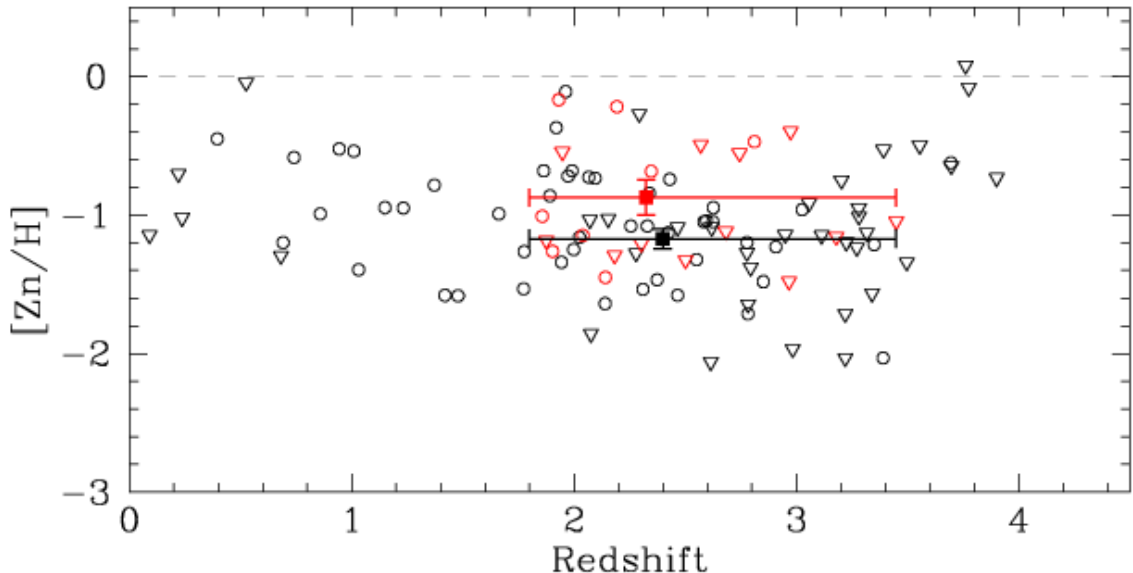}}}}
  \caption{Metallicity of DLAs in the CORALS sample (red symbols, Akerman
et al. 2005) and optically
selected samples (black symbols from the compilation of Kulkarni et al. 2005).
Open circles are detections, downward pointing triangles are upper limits.
Binned, column density weighted values (filled points, same
colour coding) are shown with bootstrap errors.  
 }\label{coralsz}
\end{figure}

In Figure \ref{coralsz} we compare our Zn abundances (Akerman et al. 2005)
with those from the compilation of Kulkarni et al. (2005) and also
show the column density weighted mean in the redshift bin $1.8 < z < 3.5$.
The difference between CORALS and the literature values depends
somewhat on the treatment of the upper limits, an issue which is
discussed more completely by Akerman et al. (2005).  In Figure
\ref{coralsz} we show the most conservative case (i.e. the
treatment which yields the largest difference between the two
samples) where the upper limits are treated as detections.
There is a marginally higher metallicity in the CORALS DLAs, but
the difference is only 0.2 dex.  

\subsection{Estimating Extinction from \bmk\ Colours}

The results described in the previous sections indicate that
the effect of dust in DLA samples is not large; within the
error bars most of the absorber statistics are in good agreement
with previous surveys.  However, as discussed in the introduction,
we are confident that \textit{some} dust is present in DLAs
based on [Zn/Cr] ratios (and may result in slightly lower number
densities in the brightest QSO samples, see \S \ref{highz}).
Based on fitting the spectral slopes of SDSS QSOs, Murphy \&
Liske (2004) have recently reported that the reddening associated
with DLA galaxies is small: \ebmv $<$ 0.02.  Michael Murphy
has presented an extension of that work at this conference
and finds that the reddening signal in the SDSS DR3 indicates
a very small \ebmv.

In order to determine the reddening in the high $z$ CORALS sample,
we have obtained near-simultaneous optical and IR photometry
of 46/66 QSOs (including 14 DLAs).  The advantage of using optical--IR
colours rather than spectral index fitting is that a much wider wavelength
baseline can be covered, potentially offering a more sensitive lever
on the dust reddening.  Although the SDSS sample offers the benefit of
a large statistical sample, it is magnitude limited, whereas CORALS
is optically complete.  The main observational challenge, however,
has been to obtain near simultaneous optical and IR photometry for the
CORALS QSOs. This
is necessary due to potentially rapid variability of radio loud
QSOs.  The data were taken with the optical (SuSI2) and IR (SofI) 
imagers on the NTT at La Silla; since these are both mounted at the
same focus, we could switch easily and quickly between them during
the night.

We have calculated normalized
\bmk\ colours which account for effects such as
redshift dependence and $B$-band flux suppression due to the DLA
\lya\ (see Ellison, Hall \& Lira 2005 for more details).  
A Kolmogorov-Smirnov test on the colours of QSOs with and without
DLAs yields only a 25\% probability that the two
colour distributions are drawn from the same parent population.
In order to quantify the amount of dust implied by this result,
we can de-redden the colours of QSOs with DLAs according to an
adopted dust recipe.  If we assume, for example, that the \ebmv\ is
fixed for every DLA, we find a most probable value of \ebmv\ = 0.02
(0.05) for SMC (Milky Way) extinction.  If we assume that
\ebmv\ is dependent on \nhi, we determine a best fit
of \ebmv\ = \nhi / $2.9 \times 10^{22}$, i.e. a reddening-to-gas
ratio some 30--50\% higher than in the SMC.

\section{Bringing It All Together}

\begin{figure}
\centerline{\rotatebox{270}{\resizebox{9cm}{!}
{\includegraphics{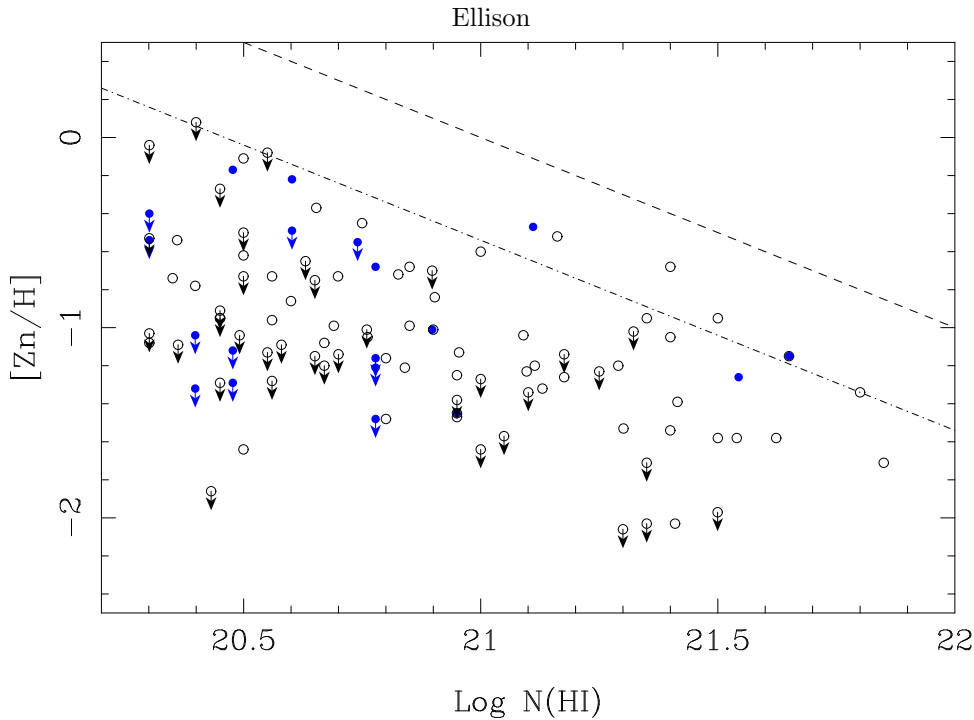}}}}
  \caption{The apparent anti-correlation between metallicity
and N(HI) in DLAs identified in optically selected QSO samples
(open points) appears to persist in CORALS (filled points).
The dashed line corresponds to the dust filter proposed by
Prantzos \& Boissier (2000), log N(HI) + [Zn/H] $<$ 21.
The dot-dashed line shows the relationship between metallicity
and N(HI) for a Galactic gas-reddening with \ebmv\ = 0.05.}\label{znh}
\end{figure}

\begin{figure}
\centerline{\rotatebox{270}{\resizebox{9cm}{!}
{\includegraphics{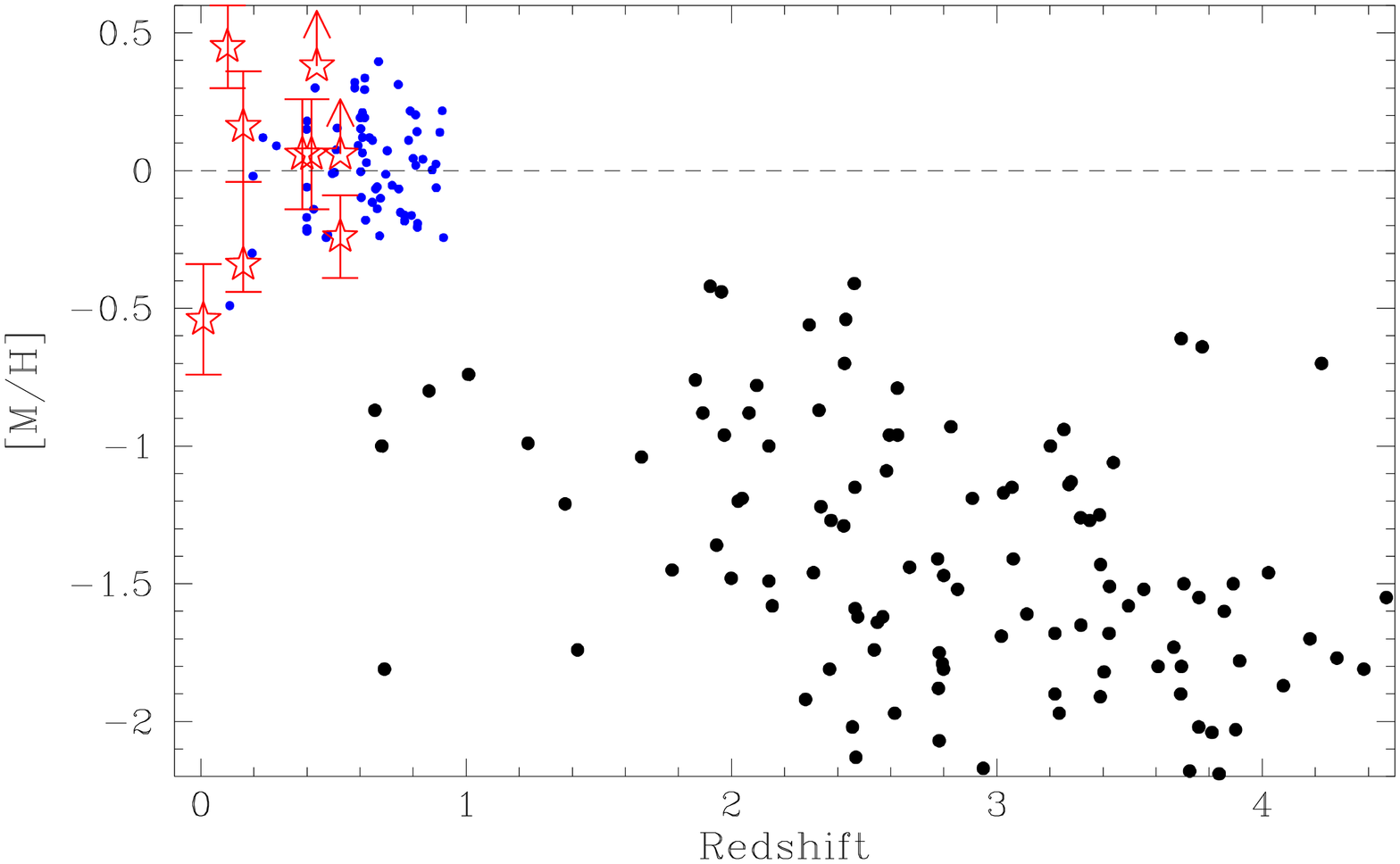}}}}
  \caption{The metallicity of DLAs (black points at $z>0.5$), emission line galaxies
(blue points at $z<1$) and absorption selected galaxies (DLAs and sub-DLAs) measured
in \textit{emission} (open stars).  DLA galaxies have $\sim$ solar
metallicities when measured in emission, compared with uniformly sub-solar
abundances when measured in absorption.}\label{elg}
\end{figure}

The presence of dust in DLAs has been well established since the 1980s
when [Zn/Cr] ratios were first shown to be super-solar.  The key
question is whether there is sufficient dust to bias our statistics,
e.g. reducing the observed number density as well as the inferred gas
and metals content.  The body of work described above has not provided
any grounds for such a concern; all evidence (so far) points to only
minor differences between magnitude limited QSO surveys and our
optically complete, radio selected one.  The lack of dependence
of $\Omega_{\rm DLA}$ and $n(z)$ on magnitude for $B>19.5$
indicates that surveys complete to this limit are unlikely to
suffer serious dust selection effects.  However, we do see evidence
that dust is present, for example the dependence of $n(z)$ and 
$\Omega_{\rm DLA}$ on $B$-band magnitude for the brightest QSOs
and the slightly redder colours
of QSOs with intervening DLAs.  For a fixed \ebmv\ in every DLA
and an SMC extinction curve (see Ellison, Hall \& Lira 2005 for other 
dust 'recipes') we determine \ebmv\ $<$ 0.05 at 99\% confidence and a best 
fit value of 0.02 for SMC extinction.   This corresponds to an 
expected observed 
frame $A_V \sim 0.25$ magnitudes, a relatively small effect.
Although the number of DLAs in the high $z$ CORALS sample is
modest ($\sim$ 20) a second survey by the UCSD group
has now almost doubled the number of radio-selected QSOs (Jorgenson
et al. in preparation)
Encouragingly, the combined samples, now with a redshift path
of $\Delta z \sim$ 100, yield a value of $\Omega_{\rm DLA}$ in good agreement
with CORALS and optically selected samples.  

The apparent anti-correlation between [Zn/H] and \nhi\ is
often attributed to dust bias, but in Figure \ref{znh} we show
that none of the CORALS DLAs (which, we recall, are not subject
to such selection effects) appear in the 'dust-forbidden' region.
However, high \nhi, high [Zn/H] values are expected to be
quite rare since these are the sightlines which presumably
pass through galaxies at the smallest impact parameters.
The simulations of Churches et al. (2004), presented by Alistair
Nelson at this meeting, indicate that $\lesssim$ 9\% (depending
upon spin parameter)
of DLAs will be in this region, so it maybe premature
to interpret the the lack of high \nhi, high metallicity
CORALS DLAs as evidence that the anti-correlation is not
caused by dust.
However, it is interesting to calculate what \ebmv\ is implied
by the envelope of DLA detections.  If we assume a Galactic
reddening to gas ratio which scales as metallicity, i.e.
\ebmv\ = \nhi\ $\times 10^{[Zn/H]} / 5.9 \times 10^{21}$, 
the empirical cut-off suggested by Prantzos \& Boissier (2000) corresponds
to quite a large reddening: \ebmv = 0.17.  On the other hand,
our reddening results, for a Milky Way extinction curve, give
\ebmv\ = 0.05.  Drawing the corresponding relation between \nhi\
and [Zn/H] on Figure \ref{znh} yields a line which
fits snugly along the detections.  This indicates that if the
anti-correlation in Figure \ref{znh} \textit{is} caused by
dust then the reddening may be small.

The CORALS results have not provided any indication that the
typically low metallicities in DLAs are due to dust bias.  Although
the low redshift CORALS sample does not have measured metallicities,
it is interesting to speculate what alternative explanations
there may be for the lack of strong evolution in [Zn/H] down to
$z \sim 0.5$.  Two contributions at this meeting, I believe, are
important clues to this outstanding question.   First, Marten Zwaan
has used 21cm maps of galaxies in the local universe, combined
with assumptions of their metallicity and abundance gradients,
to infer the $z=0$ mean weighted metallicity.  The value found
is $\sim$ 1/3 solar, indicating that we should perhaps not expect
the DLA metallicities to evolve to solar at low $z$.  Second,
Hsiao-Wen Chen has presented her results on the emission line metallicities
of galaxies associated with DLA absorbers at low $z$.
Chen's results, combined with similar work from Schulte-Ladbeck
et al. (2004) and Ellison, Kewley \& Mallen-Ornelas (2005), are
shown in Figure \ref{elg}.  We can see that when measured from
emission lines, DLAs have typically solar abundances, similar
to other bright emission line galaxies at $z<1$.  These
results indicate that sub-solar abundances in DLAs may be
due, at least in part, to substantial galaxy--QSO impact parameters.
Whereas solar metallicities in emission line galaxy samples are
weighted by luminosity, DLA measurements are weighted by HI
cross sectional area.
The tenet that DLAs are a representative probe of gas and metals
is then secure without the need to invoke grave observational biases.
Nonetheless, larger complete surveys are desirable to improve the 
statistics derived from CORALS and I think that
optically faint X-ray selected QSOs (which are being found in large
number by surveys with XMM and Chandra) provide the most promising
prospect in this direction.

\begin{acknowledgments}
I am extremely grateful to my past and present collaborators who
have contributed to various aspects of the CORALS survey:
Max Pettini, Chris Akerman, Chris Churchill, Pat Hall, Isobel Hook, 
Carole Jackson, Paulina Lira, Samantha Rix, Peter Shaver, Chuck Steidel, 
Jasper Wall and Lin Yan.  I am fortunate to have such insightful and 
productive colleagues with whom it is a joy to work.  I would
also like to thank the organisers of this meeting which has
combined a fascinating destination with a stimulating program.
\end{acknowledgments}

\end{document}